\documentclass[journal]{IEEEtran}

\usepackage{cite}
\usepackage{amsmath,amssymb,amsfonts,mathtools,bm}
\usepackage{graphicx}
\usepackage{booktabs,multirow,array}
\usepackage{url}
\usepackage{xcolor}
\usepackage[hidelinks]{hyperref}
\usepackage{subfiles}
\usepackage{comment}
\usepackage{authblk}
\usepackage{threeparttable}
\usepackage{tabularx}
\usepackage{dblfloatfix}
\usepackage{float}
\begin{document}

\title{Quantum Annealing for Dynamic Portfolio Optimization under Realistic Transaction Costs}
%\myCommentGBJ{``Hybrid Quantum-Classical Optimization for Dynamic Portfolio Strategies under Transaction Costs'', 
%``Quantum-Derived Dynamic Portfolio Optimization under Realistic Transaction Costs'', 
%o una variante similar que mantenga el foco en transaction costs sin sobreenfatizar una victoria general de quantum sobre classical methods.?}\myCommentESM{Pongo un remix y hacemos guantes en la presencial}

\author[1,2]{Escolástico Sánchez-Martínez}
\author[1]{Senaida Hernandez Santana}
\author[1]{Ventura Sarasa Laborda}
\author[1]{Pablo Serrano Molinero}
\author[2]{Guillermo Botella Juan}
\author[2]{Alberto del Barrio García}
\affil[1]{BBVA Quantum}
\affil[2]{Department of Computer Architecture and Automation, Complutense University of Madrid}

\begin{comment}
\thanks{This work was supported in part by [Funding Agency] under Grant [Grant Number].}%
\thanks{First A. Author is with [Department], [Institution], [City], [Country] (e-mail: author1@example.com).}%
\thanks{Second B. Author is with [Department], [Institution], [City], [Country] (e-mail: author2@example.com).}%
\thanks{Third C. Author is with [Department], [Institution], [City], [Country] (e-mail: author3@example.com).}%
\thanks{Corresponding author: First A. Author (e-mail: author1@example.com).}
\end{comment}

\markboth{IEEE Transactions on Quantum Engineering,~Vol.~XX, No.~XX, Month~202X}{Author \MakeLowercase{\textit{et al.}}: Quantum Beats Some Classical Investment Strategies in Business and Hardware Metrics}

\maketitle

\begin{abstract}
This paper investigates and compares quantum and classical investment strategies for portfolio construction under realistic trading and allocation constraints. The study considers a multi-period portfolio re-balancing setting in which asset weights are subject to lower and upper bounds, class-level exposure restrictions, and turnover limitations. On the quantum side, the portfolio selection problem is reformulated as a quadratic unconstrained binary optimization (QUBO) model through a finite binary encoding of asset weights, and subsequently embedded into a constrained quadratic model (CQM) solved by a hybrid quantum-classical optimization workflow. On the classical side, benchmark allocation strategies are introduced to provide an economically meaningful reference for performance assessment. The proposed framework jointly evaluates portfolio efficiency from two complementary perspectives: business performance, measured through return, volatility, and a measure of the trade-off between them, the Sharpe ratio; and computational performance, assessed through solver structure, constraint handling, and hardware execution characteristics. The resulting comparison provides a rigorous basis for understanding the practical role of quantum annealing in constrained portfolio optimization, while clarifying both its modeling advantages and its current implementation limitations relative to established classical approaches.
\end{abstract}


% =================================================
\section{Introduction}

Dynamic Portfolio Optimization (DPO) is a central challenge in quantitative finance, where the objective is to determine the optimal trading strategies over a time period, balancing return and risk. In realistic settings, the inclusion of transaction costs and market impact significantly increases the complexity of the problem, often rendering classical approaches computationally intractable. In this context, \textbf{quantum and quantum-inspired methods} have emerged as promising alternatives, as they enable efficient formulations of multi-period portfolio optimization problems through Quadratic Unconstrained Binary Optimization (QUBO) representations and hybrid optimization frameworks\cite{dynamic-portfolio-optimization,Rosenberg:2015fem,DeLeon:2025hzl,Nodar:2024gwl}.

Despite the theoretical capabilities, it remains unclear under which practical conditions quantum methods provide a measurable advantage over classical approaches. In particular, \textbf{the role of transaction costs as a determining factor in this potential advantage has not been systematically studied. In this work, we address this gap by evaluating the performance of quantum optimization methods across a range of transaction cost regimes.}

We benchmark quantum-derived trading strategies against standard classical approaches, including buy-and-hold and Conditional Value-at-Risk (CVaR) optimization \cite{Rockafellar2000OptimizationOC}. By doing so, we identify the transaction cost regimes in which quantum methods outperform classical alternatives in realistic multi-period portfolio settings, considering a 21-asset portfolio over a long-term investment horizon.

The integration of quantum computing into quantitative finance has attracted significant attention from major global financial institutions aiming to address computationally intractable optimization and risk management problems. Recent studies have explored applications ranging from portfolio optimization to risk modeling and machine learning, including works by JPMorgan Chase \cite{Yalovetzky2024, Acharya:2024sxl}, Vanguard \cite{Agliardi:2025rxn}, and Crédit Agricole \cite{Leclerc:2022iki}, highlighting both the potential and the current limitations of quantum approaches.

In particular, these efforts emphasize the challenges associated with achieving a practical quantum advantage, including scalability constraints, hardware limitations, and the need for further empirical validation. A critical perspective is provided by Goldman Sachs \cite{PRXQuantum.4.040325}, which shows that significant algorithmic improvements are still required before quantum methods can outperform classical approaches in realistic settings. This highlights the need for systematic evaluations under realistic conditions, which is precisely the focus of this work.

Previous work has explored the application of quantum and quantum-inspired methods to dynamic portfolio optimization using real-world datasets \cite{dynamic-portfolio-optimization,Rosenberg:2015fem,DeLeon:2025hzl,Nodar:2024gwl}, including approaches based on quantum annealers, variational quantum eigensolvers and tensor networks. These studies demonstrate the feasibility of scaling such methods to portfolios with order of tens of assets (30–50 assets) achieving competitive performance with respect to classical solvers.

Building on this line of research, we consider a hybrid workflow composed of two main stages: The first stage involves dimensionality reduction through clustering techniques, which can be used to adapt large-scale financial problems to the constraints of current quantum hardware. The second stage focuses on the optimization of trading trajectories, maximizing expected  returns (including transaction costs) while minimizing risk.

In the present work, we focus specifically on the optimization stage.%, which represents the core of this analysis. 
~We present the optimal trading trajectory of a realistic portfolio investment strategy by maximizing expected returns while minimizing financial risk. The investment universe is formed by 21 global assets (equities, government bonds, credit, and forex within Europe, America, and Asia), with constraints in individual and asset class punctual and incremental weights, with an average fee of 7 basis points (0.07\%) of the transaction volume. We construct three strategies, one of them quantum-derived and the other two classical ones.

According to the experimental results, after comparing the ex-post performance of the three investment strategies in terms of return and risk, \textbf{when the level of fees reaches 35 basis points (0.35\%) of the transaction amount, the quantum'-derived investment strategy starts beating the classical ones.} The market impact in our portfolio, with 7 basis points, is not relevant because the assets in our exercise are very liquid.
The rest of the paper is organized as follows:
\begin{itemize}
    \item \textbf{Section II} provides the theoretical background, specifically focusing on Modern Portfolio Theory and its classical and quantum possible implementations.
    \item \textbf{Section III} defines the problem by detailing the investment universe across multiple rebalancing dates and outlining the layers of constraints applied to the portfolio.
    \item \textbf{Section IV} outlines the methodology, describing the classical benchmark approaches and the Quadratic Unconstrained Binary Optimization (QUBO) model.
    \item \textbf{Section V} presents the empirical results, evaluating the economic and hardware quality, as well as the performance metrics of the resulting portfolios.
    \item \textbf{Section VI} concludes the study by summarizing the structured comparison between the quantum and classical strategies.
    \item \textbf{Section VII} proposes directions for future work.
    \item \textbf{Appendix A} supplies additional mathematical details, such as the full derivation of the QUBO coefficients and the binary encoding mapping.
\end{itemize}
% =================================================
\section{Theoretical Background}
\subsection{Classical Approach}

Modern Portfolio Theory (MPT) provides the foundational mathematical framework for portfolio construction. Introduced by Harry Markowitz in 1952, the core objective of MPT is to select an allocation of assets within an investment universe that maximizes expected returns while simultaneously minimizing financial risk \cite{portfolio-selection-1952}. Traditionally, the risk is quantified as the variance or standard deviation (volatility) of the portfolio's returns. 
Let \(w=(w_1,\dots,w_N)^\top\) denote the portfolio weights, adding to 1, \(\mathbf{1}^\top w=1\), let \(\mathcal{R}=(\mathcal{R}_1,\dots,\mathcal{R}_N)^\top\) denote the random asset-return vector where $\mathcal{R}_i$ denotes the arithmetic return of asset $i$, let \(\mu=\mathbb{E}[\mathcal{R}]\) denote the expected return vector, and let \(\Sigma=\operatorname{Cov}(\mathcal{R})\in\mathbb{R}^{N\times N}\) denote the covariance matrix. The return of the portfolio, without taking into account transaction cost, will be $\mathcal{R}_{port}=w ^\top \mathcal{R} $, with expected value $\mathbb{E}[\mathcal{R}_{port}]=w ^\top \mu\ $ and the risk will be the variance of the portfolio return $w^\top \Sigma w$ or its square root (volatility).

For a target return \(\bar r\),  the classical Markowitz problem can then be written as
\begin{equation}
\min_{w\in\mathcal{W}} \; w^\top \Sigma w
\qquad \text{subject to} \qquad
w^\top \mu=\bar r, \qquad \mathbf{1}^\top w=1.
\end{equation}
where \(\mathcal{W}\) denotes the space of weights satisfying any additional feasibility constraints. 
Mixing risk and reward, the mainstream metric is the Sharpe ratio, which quantifies the amount of return generated per unit of risk. A higher Sharpe ratio indicates substantial returns relative to the assumed risk, whereas a ratio near zero or negative indicates poor performance and losses \cite{sharpe-ratio-1994}. If the portfolio return is \(\mathcal{R}_{port}=w^\top \mathcal{R}\) and the risk-free rate is \(r_f\), then the Sharpe ratio is
\begin{equation}
\operatorname{SR}(w)=\frac{\mathbb{E}[\mathcal{R}_{port}]-r_f}{\sqrt{\operatorname{Var}(\mathcal{R}_{port})}}
=
\frac{w^\top\mu-r_f}{\sqrt{w^\top\Sigma w}}.
\label{eq:Sharpe}
\end{equation}

 Thanks to evaluating the covariances between different pairs of assets, MPT demonstrates the fundamental benefits of diversification to reduce overall portfolio risk \cite{0eaa333e-d012-3921-bd03-26ecd9e6a037}. A central concept in this framework is the efficient frontier, which represents the location of the optimal portfolios offering the highest possible expected return for any specified level of market risk \cite{efficient-portfolio-frontier-1972}. The efficient frontier is obtained by solving, for varying target returns \(\bar r\),
\begin{equation}
\min_{w\in\mathcal{W}} \; w^\top\Sigma w
\qquad \text{subject to} \qquad
w^\top\mu=\bar r,
\qquad
\mathbf{1}^\top w=1,
\end{equation}
or equivalently, for target volatility \(\bar\sigma\),
\begin{equation}
\max_{w\in\mathcal{W}} \; w^\top\mu
\qquad \text{subject to} \qquad
w^\top\Sigma w \le \bar\sigma^2,
\qquad
\mathbf{1}^\top w=1.
\end{equation}
To some extent, we can summarize writing
\begin{equation}
\max_{w\in\mathcal{W}} \; w^\top \mu - \gamma\, w^\top \Sigma w
\qquad \text{subject to} \qquad \mathbf{1}^\top w=1,
\label{eq:mean-variance}
\end{equation}
where \(\gamma>0\) is a risk-aversion parameter.
While variance or volatility are the standard risk metrics in classical MPT, modern risk management emphasizes the need to evaluate extreme downside losses. Two prominent measures are used for this purpose:
\begin{itemize}
    \item \textbf{Value-at-Risk (VaR)}: For a specified confidence level (or significance level) \(p\in(0,1)\), $\operatorname{VaR}_p$ is the 
$p*100$ percentile of the portfolio loss distribution (assumed as continuous) \cite{Jorion}. Let the loss random variable be \(L=-\mathcal{R}_{port}\). Then
%$(1-p)* 100$ percentile of the portfolio loss distribution (assumed as continuous). Let the loss random variable be \(L=-\mathcal{R}_{port}\). Then
\begin{equation}
\mathbb{P}(L\le \operatorname{VaR}_{p}(L))=p%1-p.
\end{equation}

Despite its popularity, VaR has undesirable mathematical characteristics, such as a lack of subadditivity and convexity. In particular, VaR may violate the subadditivity inequality
\begin{equation}
\operatorname{VaR}(L_1+L_2)\le \operatorname{VaR}(L_1)+\operatorname{VaR}(L_2),
\end{equation}
which means that a combined portfolio can be classified as riskier than its individual components. It can also exhibit multiple local extrema, making it difficult to optimize \cite{coherent-measures-of-risk-1999}.

\item \textbf{Conditional Value-at-Risk (CVaR)}: Also known as Mean Excess Loss or Tail VaR, CVaR measures the conditional expectation of losses that exceed the VaR threshold. It is defined as
\begin{equation}
\operatorname{CVaR}_{p}(L)
=
\mathbb{E}\!\left[L \mid L\ge \operatorname{VaR}_{p}(L)\right].
\end{equation}
An equivalent optimization-friendly representation \cite{optimization-cvar-2000} is
\begin{equation}
\operatorname{CVaR}_{p}(L)
=
\min_{\alpha\in\mathbb{R}}
\left\{
\alpha+\frac{1}{1-p}\,\mathbb{E}\!\left[(L-\alpha)_+\right]
\right\}~,
\end{equation}
provided that \((x)_+=\max\{x,0\}.\)

Unlike VaR, CVaR is a coherent risk measure that is positively homogeneous, convex, and mathematically superior for optimization \cite{optimization-cvar-2000}. Since CVaR also satisfies by definition
\begin{equation}
\operatorname{VaR}_{p}(L)\le \operatorname{CVaR}_{p}(L),
\end{equation}

portfolios with low CVaR inherently must have controlled tail losses at the same confidence level, so minimizing CVaR is considered a more consistent and effective method for hedging against extreme financial risks \cite{b91e0438-d4d3-327f-bb28-37474d398566}.

\end{itemize}
Within this framework, the covariance matrix of asset returns is the basic input for calculating risk measures. While these approaches provide a mathematically elegant and tractable foundation, they rely on simplifying assumptions—most notably, that investors only care about the first two moments of return distributions (in Gaussian distribution these two moments define the whole random variable of return), and that inputs such as expected returns and covariances are known or can be reliably estimated \cite{approximating-expected-utility-1979}. The element $(i,j)$ of the covariance matrix is defined as
\begin{equation}
\Sigma_{ij}
=
\operatorname{Cov}(\mathcal{R}_i,\mathcal{R}_j)
=
\mathbb{E}\!\left[(\mathcal{R}_i-\mu_i)(\mathcal{R}_j-\mu_j)\right],
\end{equation}
with
\begin{equation}
\mu_i=\mathbb{E}[\mathcal{R}_i].
\end{equation}
If we assume Gaussian
\begin{equation}
\mathcal{R} \sim \mathcal{N}(\mu,\Sigma),
\end{equation}
the pair \((\mu,\Sigma)\) fully characterizes the return distribution. In practice, these quantities are estimated from data, for example by
\begin{equation}
\hat\mu_i=\frac{1}{T}\sum_{t=1}^{T} \mathcal{R}_{t,i},
\qquad
\hat\Sigma_{ij}
=
\frac{1}{T-1}
\sum_{t=1}^{T}
(\mathcal{R}_{t,i}-\hat\mu_i)(\mathcal{R}_{t,j}-\hat\mu_j).
\end{equation}

However, real-world financial markets exhibit structural complexity that challenges these assumptions. Asset returns are often non-Gaussian, correlations are time-varying \cite{empirical-properties-of-asset-returns-2001}, and estimation errors can significantly distort optimal portfolios: even minor changes in the input data can cause the efficient frontier to shift considerably \cite{markowitz-optimization-enigma-1989}. Classical optimization methods also face computational limitations: the dimensionality of the covariance matrix grows quadratically with the number of assets, making large-scale portfolio optimization increasingly difficult. 

\subsection{Quantum Approach}

Quantum computing offers a fundamentally different computational paradigm that may help address some of these limitations in certain settings. At its core, quantum computation exploits superposition and entanglement to process information in high-dimensional spaces. This could allow certain optimization and sampling problems—central to portfolio construction—to be addressed within alternative optimization frameworks that may offer advantages in specific cases \cite{quantum-algorithms-overview-2016}. A single qubit is represented by
\begin{equation}
|\psi\rangle=c_0 |0\rangle+c_1 |1\rangle,
\qquad
|c_0|^2+|c_1|^2=1,
\end{equation}
and an \(n\)-qubit register lives in a Hilbert space of dimension \(2^n\), with general state
\begin{equation}
|\Psi\rangle
=
\sum_{z\in\{0,1\}^n} c_z |z\rangle,
\qquad
\sum_{z\in\{0,1\}^n} |c_z|^2=1.
\end{equation}
Entanglement appears when the joint state cannot be factorized as
\begin{equation}
|\Psi\rangle \neq \bigotimes_{k=1}^{n} |\psi_k\rangle.
\end{equation}

For instance, quantum annealing and quantum algorithms can be used to solve combinatorial portfolio optimization problems that become computationally challenging for classical methods \cite{PhysRevLett.134.160601}, particularly when discrete decision variables and added penalty functions (e.g., cardinality constraints or transaction costs) are introduced \cite{reverse-quantum-annealing-portfolio-2019}. 

Nevertheless, the quantum advantage remains largely theoretical at present. 
Current hardware limitations—such as noise, limited qubit counts, and short coherence times—restrict practical implementation \cite{nisq-era-and-beyond-2018}. Moreover, the mapping between financial problems and quantum circuits introduces additional overhead that may offset theoretical gains in near-term devices. If \(N\) assets are encoded with \(Q\) binary variables each, then the binary problem already contains
\begin{equation}\nonumber
n_{\mathrm{bin}}=NQ
\end{equation}
decision variables, while the number of pairwise quadratic couplings can scale as
\begin{equation}\nonumber
O(n_{\mathrm{bin}}^2).
\end{equation}
For gate-based hardware, useful circuit depth \(d\) is further constrained by coherence times, roughly requiring
\begin{equation}\nonumber
d\,t_g \ll T_2,
\end{equation}
where \(t_g\) is the characteristic gate time and \(T_2\) is the coherence time.

In summary, while classical investment strategies rooted in Modern Portfolio Theory provide a robust conceptual baseline, they face both empirical and computational constraints. Quantum computing introduces a promising avenue for addressing these challenges by enabling more efficient optimization, richer modeling of dependencies, and improved handling of uncertainty. The extent to which this potential translates into practical financial advantage remains an open and active area of research \cite{quantum-computing-for-finance-2023}. 

Formally, two related but distinct questions remain open: i) whether quantum-derived methods can improve portfolio performance metrics, such as the Sharpe ratio in equation \eqref{eq:Sharpe}, and ii) whether they can also provide a computational advantage over classical methods for financially relevant problem instances
\begin{equation}\nonumber
T_Q(n) < T_{C}(n),
\end{equation}
where \(T_Q(n)\) and \(T_{C}(n)\) denote the effective quantum and classical computational costs for a problem of size \(n\), especially for large values of $n$.
% =================================================

\section{Problem definition}

\subsection{Investment Universe}
An investment portfolio is a combination of financial assets, which have a value that depends on time (usually market price or fair value). 
When the portfolio is composed by $n_{t,i}$ units of asset $i$ in the set $\{1,\ldots,N\}$, with values $P_{t,i}$ in instant $t$, its value will be:
\begin{equation}
V_t=\sum^{N}_{i=1}n_{t,i}\,P_{t,i}
\end{equation}
The arithmetic return of the portfolio from instant $0$ to instant $t$ (typically referred to simply as the portfolio return) is the difference of the portfolio values divided by the initial value, assuming a buy-and-hold policy (i.e. $n_{t,i}=n_{0,i}=n_i$ is fixed).
Therefore, the portfolio return\cite{ang2021portfolio} will be:

%When the portfolio is composed by $n_{i}$ units of asset $i$ in the set $\{1,\ldots,N\}$, with values $P_{t,i}$ in instant $t$, its return will be:

% \begin{equation}\label{eq:portfolio-opt}
% \begin{split}
% R_{p}(0,t)&=\frac{V_{t}-V_{0}}{V_{0}}=\frac{\sum^{N}_{i=1}n_{i}P_{t,i}-\sum^{N}_{i=1}n_{i}P_{i}_{0}}{\sum^{N}_{i=1}n_{i}P_{i}_{0}}\\
%    & =\frac{\sum^{N}_{i=1}n_{i}P_{i}_{0}\cdot\frac{P_{t,i}-P_{i}_{0}}{P_{i}_{0}}}{\sum^{N}_{i=1}n_{i}P_{i}_{0}}=\sum^{N}_{i=1}w_{i}_{0}\cdot R_{i}(0,T)\\
%    &=W^{T}\cdot R \ \ \ with\\
%    \  &\ \ \ \ \ w_{t,i}:=\frac{n_{i}P_{t,i}}{\sum^{N}_{j=1}W_{t,j}}.\\ \ \  &R_{i}(t_{1},t_{2}):=\frac{P_{i}(t_{2})-P_{i}(t_{1})}{P_{i}(t_{1})}\ \ \ and \ \ \  \sum^{N}_{i=1}w_{t,i}=1.
% \end{split}
% \end{equation}

\begin{align}
\mathcal{R}_{port}(0,t)
&= \frac{V_{t}-V_{0}}{V_{0}} \\
&= \frac{\sum_{i=1}^{N} n_{i} P_{t,i} - \sum_{i=1}^{N} n_{i} P_{0,i}}
         {\sum_{i=1}^{N} n_{i} P_{0,i}} \\
&= \frac{\sum_{i=1}^{N} n_{i} P_{0,i}\,\frac{P_{t,i}-P_{0,i}}{P_{0,i}}}
         {\sum_{i=1}^{N} n_{i} P_{0,i}} \\
&= \sum_{i=1}^{N} w_{0,i}\, \mathcal{R}_i(0,t) = w^\top \mathcal{R}
\label{eq:portfolio-opt}
\end{align}

%\vspace{-0.9em}
where the arithmetic return $\mathcal{R}_i$ of asset $i$ between times $t_1$ and $t_2$ is defined as
\begin{equation}\label{eq:weight}
\mathcal{R}_i(t_1,t_2):=\frac{P_{t_2,i}-P_{t_1,i}}{P_{t_1,i}},
\end{equation}
and its weight is given by
\begin{equation}\label{eq:weight}
\begin{array}{cc}
\displaystyle
w_{t,i}:=\frac{n_{t,i}P_{t,i}}{V_t}.
\end{array}
\end{equation}

%\vspace{-1em}

%\begin{equation}
%\sum_{i=1}^N w_{t,i}=1.
%\end{equation}

In addition, we model proportional transaction fees \cite{lobo2007portfolio}. Let $f_i$ denote the percentage fee on effective amount for trading asset $i$, and let $Tr_{t,i}=n_{t+1,i}-n_{t,i}$ denote the number of units traded at re-balancing interval $[t,t+1]$. The total transaction cost in that re-balancing interval is (assuming, for simplicity that transaction prices are $P_{t+1,i})$:

\begin{equation}
    \mathrm{TC}_t = \sum_{i=1}^{N} f_i\, |Tr_{t,i}|\, P_{t+1,i}.
\end{equation}
Most of the times, for technical purposes, with little loss of accuracy, as can be seen in \cite{TFMEsco}, instead of working with the simple arithmetic returns $\mathcal{R}_i(t_{1},t_{2})$, we will work with logarithmic returns, defined as 
\begin{equation}
R_i(t_{1},t_{2})=\ln\frac{P_{t_2,i}}{P_{t_1,i}} .
\end{equation}

Consider an investment universe with \(N\) assets observed over \(T\) rebalancing dates. Let \(t=0,\dots,T-1\) index time and \(i=1,\dots,N\) index assets. For each period \(t\), let \(R_{t,i}\) denote the expected or realized log-return input associated with asset \(i\) (if $i=port$ we are talking about the whole portfolio) from time $t$ to time $t+1$, and let $\Sigma_t \in \mathbb{R} ^{N \times N}$denote the covariance matrix used for risk penalization. For convenience, from now on, we will be working with log returns along all the different implementations. 

Assets are partitioned into a finite set of categories or asset classes. Let \(\mathcal{A}_k \subseteq \{1,\dots,N\}\) denote the set of assets belonging to class \(k\), for \(k=1,\dots,K_c\) . This structure is introduced to capture strategic allocation rules at the class level, such as minimum and maximum exposure \cite{abate2022constraints,kolm2014sixty}.
The investment universe comprises the following asset classes: \textbf{equities}, represented by MSCI indices for the United States, Europe, Japan, and Emerging Markets; \textbf{government bonds}, including short-, medium-, and long-term sovereign bonds from the United States and Europe, as well as hard-currency Emerging Markets sovereign bonds; \textbf{credit}, including U.S., European, and pan-European corporate bonds; \textbf{foreign exchange (FX)}, consisting of EUR, JPY, CAD, AUD, and MXN, all quoted against the USD; and \textbf{cash}, represented by the euro (EUR). More details about the investment universe in Appendix \ref{Assets}.

In this problem, we analyze 21 assets grouped into $K_c=5$ asset classes, using data from 418 biweekly periods spanning from January $6^{th}$, 2003 to December $31^{st}$, 2018. At each date \(t\), the decision variable to be optimized is the portfolio weight vector
\begin{equation}\label{eq:weights}
    w_t = (w_{t,1},\dots,w_{t,N})^\top \in \mathbb{R}^{N},
\end{equation}
where $w_{t,i}$ corresponds to the portfolio weight of asset $i$. Moreover, we consider an average of transaction fees of $f=0.07\%=0.0007$.
\textbf{The objective is to maximize the return of our portfolio, taking into account transaction cost, and minimize the risk of the return (its variance). That is, } 
%\textcolor{red}{
\begin{equation}
\max_{w} \quad \mathbb{E}[R_{port}] - \mathbf{Vol}^2(R_{port}) - TC(w)~~\text{s.t}~\mathcal{W}~\text{are satisfied} . 
\end{equation}
%}

Being $\mathbf{Vol}^2(R_{port})$ the variance of the portfolio returns.

\subsection{Constraints}\label{Constraints}
The optimization problem is subject to a set of feasibility constraints $\mathcal{W}$. These constraints fall in three categories: asset-level constraints, which impose lower and upper bounds on individual asset weights; asset-class constraints, which define minimum and maximum exposures for each asset class; and trading constraints, which limit the quantity that can be bought or sold between rebalancing dates \cite{kolm2014sixty,abate2022constraints}.

First, each asset weight is bounded:
\begin{equation}
    l_i \le w_{t,i} \le u_i,
    \hspace{0.4cm} i=1,\dots,N,\; \hspace{0.4cm} t=0,\dots,T-1.
\end{equation}
Second, the portfolio must satisfy the budget constraint. Since FX assets can have both long (buying) and short (selling) positions, this category of assets are excluded of this constraint.
\begin{equation}
    \sum_{i\notin FX} w_{t,i} = 1,
    \qquad t=0,\dots,T-1.
\end{equation}

Third, class-level exposures must remain within admissible bounds 
\begin{equation}
    L_k \le \sum_{i \in \mathcal{A}_k} w_{t,i} \le U_k,
    \hspace{0.2cm} k=1,\dots,K_c,\; t=0,\dots,T-1.
\end{equation}

Fourth, the re-balancing process is constrained through turnover limits.
The monetary turnover of asset $i$ form $t$ to $t+1$ is exactly $Tr_{t,i}*P_{t+1,i}$. Considered as a percentage of the final portfolio value, the percentage turnover will be $w_{t+1,i}-w_{t,i}\frac{V_t}{V_{t+1}}\frac{P_{t+1,i}}{P_{t,i}}.$
Let \(\tau_i\) denote the admissible percentage turnover bound for asset \(i\). Then the weight assigned to asset \(i\) at time \(t\) must remain sufficiently close to the drifted previous allocation:
%\begin{equation}
%    -\tau_i \le w_{t+1,i} - e^{\mu_{t,i}} w_{t,i} \le \tau_i.
%\end{equation}
\begin{equation}
    -\tau_i \le w_{t,i} - w_{t,i}^{\text{drift}} \le \tau_i,
\end{equation}
where the drift weight $w_{t,i}^{\text{drift}}$ is the resulting weight after price evolution from $t-1$ to $t$ and before rebalancing, given by:
\begin{equation}
\label{weight_drift}
w_{t,i}^{\text{drift}} =w_{t-1,i} e^{R_{t,i}-R_{t,{port}} } 
\end{equation}
as seen above.

Similarly, class-level turnover may be restricted through bounds \(\Theta_k\):
\begin{equation}
    -\Theta_k \le
    \sum_{i \in \mathcal{A}_k} w_{t+1,i}
    -
    \sum_{i \in \mathcal{A}_k} w_{t,i}^{\text{drift}}
    \le \Theta_k.
\end{equation}

These constraints were advised by business experts and define a practitioners' way of portfolio construction in which the feasible set is shaped not only by static exposure limits, but also by the path-dependent mechanics of portfolio re-balancing.
For additional details on the investment universe and the specific constraint set used in this case study, see Appendix~\ref{Assets}.

\section{Methodology}

This section describes the benchmark strategies used in the empirical comparison and the quantum formulation adopted for dynamic portfolio optimization. The classical side includes a buy-and-hold portfolio and a downside-risk benchmark based on Conditional Value-at-Risk (CVaR). The quantum side follows the binary encoding and quadratic optimization framework developed for multi-period portfolio rebalancing in \cite{dynamic-portfolio-optimization,Rosenberg:2015fem}.

\subsection{Classical Benchmarks}

The purpose of the classical benchmarks is to provide economically interpretable reference points against which the quantum strategy can be evaluated. The first benchmark isolates the value of maintaining a fixed strategic allocation without dynamic optimization, whereas the second introduces an optimization-based classical allocation rule focused on downside-risk control.

\subsubsection{Buy-and-Hold Strategy}

% The buy-and-hold benchmark represents the simplest admissible investment policy in the multi-period setting. Despite its simplicity, buy-and-hold constitutes a widely adopted investment strategy in practice, particularly in long-term asset allocation and passive portfolio management, where minimizing turnover and transaction costs is a primary objective \cite{malkiel1973random,bogle1999common}.

% An initial feasible allocation $w_0^{BH} \in \mathcal{W}$ is selected at the first decision date and then kept throughout the investment horizon. 

The buy-and-hold benchmark represents the simplest admissible reference policy in the multi-period setting. In this work, buy-and-hold is implemented as a fixed strategic allocation benchmark with periodic rebalancing. That is, the initial allocation $w^{BH}_0 \in W$ defines the target portfolio weights, and at each rebalancing date the portfolio is brought back to these target weights. This convention makes the benchmark comparable with the active CVaR and quantum strategies, since all three strategies are evaluated under the same rebalancing frequency, transaction-cost framework, and business constraints.

Despite its simplicity, this fixed-allocation benchmark constitutes a widely adopted reference strategy in long-term asset allocation and passive portfolio management, where preserving a stable strategic allocation and controlling turnover are primary objectives \cite{malkiel1973random,bogle1999common}.

Between two consecutive rebalancing dates, portfolio weights evolve endogenously due to market returns. Let $R_{t,i}$ denote the realized log return of asset $i$ between dates $t-1$ and $t$, and let $R^{BH}_{t,port}$ denote the corresponding log return of the buy-and-hold portfolio over the same period. Before rebalancing, remind that the drifted weight of asset $i$ is given by Eq.~\eqref{weight_drift}.

The portfolio return \(r^{BH}_{t,p}\) acts as the normalization term ensuring that the drifted weights sum to one. At each rebalancing date, the portfolio is reset to the target allocation,
\begin{equation}
w^{BH}_{t}=w^{BH}_{0}.
\end{equation}
Therefore, \(w^{drift}_{t,i}\) denotes the weight of asset  $i$ before rebalancing, whereas $w^{BH}_{t,i}$ denotes the implemented benchmark weight after rebalancing.

% Between two consecutive rebalancing dates, portfolio weights evolve endogenously due to market returns. Let $r_t \in \mathbb{R}^N$ denote the vector of realized log returns between dates $t-1$ and $t$. Under the self-financing condition, the portfolio weights evolve as
% \myCommentSHS{comprobar formula, equivalente a $n_{i, t-1}*P_{i, t-1}*(1+r{i,t})/(V_{t-1}*(1+R_{t}))/(\sum_i n_{i,t-1}*P_{i,t-1}*(1+r_{i,t})/(V_{t-1}*(1+R_{t}))=n_{i, t-1}*P_{i, t}/(\sum_i n_{i,t-1}*P_{i,t})). OK$}

% \begin{equation}
% w_t^{BH} = 
% \frac{w_{t-1}^{BH} \odot (1 + r_t)}
% {1^\top \left( w_{t-1}^{BH} \odot (1 + r_t) \right)},
% \end{equation}
%that is, this is the value of $w_t^{BH}$ before rebalancing 

This benchmark should therefore be interpreted as a rebalanced buy-and-hold strategy, or equivalently as a fixed-allocation rebalanced benchmark, rather than as a pure buy-and-hold policy. It does not solve a dynamic optimization problem and does not update the target allocation using new expected-return or risk estimates. Its only active component is the mechanical rebalancing required to restore the initial strategic allocation and to keep the portfolio within the same business constraints imposed on the CVaR and quantum strategies.

As such, it provides a realistic and comparable baseline for assessing the value added by dynamic optimization. The comparison with the CVaR and quantum strategies therefore isolates the incremental effect of optimizing the allocation over time, rather than the effect of merely enforcing the common rebalancing schedule and operational constraints.

\subsubsection{CVaR-Based Classical Optimization}

The second classical baseline is an active strategy that, at every
rebalancing date, reallocates capital by solving a Conditional
Value-at-Risk (CVaR) optimization problem in the
Rockafellar--Uryasev linear formulation~\cite{Rockafellar2000OptimizationOC}.
Following standard practice in the risk-management literature, the
risk aversion of the strategy is not encoded as a convex combination
of mean and CVaR in the objective; instead, the CVaR is minimized subject to a parametric lower bound on the portfolio average return, $R_{\min}$.
%Sweeping $R_{\min}$ across its admissible range traces the entire CVaR efficient frontier, from which a single allocation is selected at each date.

\paragraph{Optimization problem}
Let $\mathbf{R}\in\mathbb{R}^{S\times N}$ be a matrix of $S$
return scenarios over the $N$ assets and let $w_{t-1}$ denote the
portfolio held at the previous rebalancing date. At date $t$ we
solve 
\begin{equation}
\begin{aligned}
  \min_{\textbf{w}, \alpha, u}\; \alpha + \frac{1}{S(1-p)}\sum_{s=1}^{S} \eta_s,  \\
  \text{s.t.}\quad &
  \eta_s\ge 0,\quad s=1,\dots,S, \\
  & \eta_s \;\ge\; -w^T\,R_{s} - \alpha,\quad \\
  & \frac{1}{S}\sum_{s=1}^{S}w^T\,R_{s} \;\ge\; R_{\min}, \\
  &w \in \mathcal{W},
\end{aligned}
\label{eq:cvar-model}
\end{equation}

where $p\in[0,1)$ is the CVaR confidence level (set to $p
=0.95$
in all experiments), $\alpha$ is the auxiliary variable that, at the optimum, coincides with the Value-at-Risk at level $p$, and $\eta_s$ are the standard slack variables that linearize the tail expectation. The set $\mathcal{W}$ collects all linear constraints defining the admissible portfolios---budget, box, and group constraints on individual assets and asset classes. Because problem \eqref{eq:cvar-model} is formulated as a linear program, it can be solved to optimality using a deterministic Linear Programming (LP) solver.

In this exercise, we obtain the  portfolio $w_t^{CVaR}$ by minimizing the CVaR with no minimum-return constraint ($R_{\min}\!=\!-\infty$).
%\myCommentSHS{i do not understand why this is CVaR, should not be minimum? $w^{CVaR}_{\text{min}}$}
The optimization problem in Eq.~\eqref{eq:cvar-model} is performed sequentially over time in a rolling-horizon fashion, that is, the period under consideration shifts forward.
At each rebalancing date $t$, the portfolio is updated by solving the CVaR optimization problem using the previously implemented allocation after evolution $w_{t-1}$ as the initial state.

This induces a path dependency through turnover constraints,  effectively resulting in a rollout policy where decisions are conditionally optimal  given past allocations but not globally optimal over the full horizon.

\paragraph{Scenario generation}
The optimization problem in Eq.~\eqref{eq:cvar-model} is solved using scenarios generated by a classical Parametric Gaussian Monte Carlo model, calibrated over a backward estimation window of ($L_{\text{window}}=35$) weeks ending at each rebalancing date. Over this window, the mean vector $\hat\mu_t$ and covariance matrix $\hat\Sigma_t$ of the  log-returns are estimated on the backward window $L_{window}$, and $S$ scenarios are drawn from $\mathcal{N}(\hat\mu_t,\hat\Sigma_t)$. This generator provides a smooth, fully parametric description of the joint distribution at the cost of imposing Gaussian tails.

\paragraph{Annualization convention}
Throughout the benchmark, expected returns are annualized by linear
scaling with $D=252$ trading days per year, while the CVaR is
annualized by combining a linear projection of the mean with a
square-root-of-time scaling of the tail term, namely
\begin{equation}
  \mathrm{CVaR}^{\mathrm{ann}}
  \;=\;
  \sqrt{D}\,\mathrm{CVaR}_d \;-\; (D-\sqrt{D})\,\bar r_d,
\end{equation}
with $\bar r_d$ and $\mathrm{CVaR}_d$ the daily mean return and daily CVaR of the portfolio, respectively. This annualization is an approximation that assumes i.i.d. Gaussian distribution and combines linear scaling of the mean and square-root-of-time scaling of the tail component, as no closed-form scaling rule exists for CVaR \cite{DanielssonZigrand2006}. More details in Annex \ref{app:cvarannaulization}.

The Sharpe ratio is annualized by multiplying the daily volatility by $\sqrt{D}$, and a zero risk-free rate is assumed in all reported figures.

\subsection{Quantum Approach}

The quantum methodology is based on the dynamic portfolio optimization framework proposed in \cite{dynamic-portfolio-optimization}, where the multi-period rebalancing problem is reformulated as a quadratic unconstrained binary optimization (QUBO) model compatible with quantum annealing (\ref{eq:mean-variance}). In practice, however, the presence of constraints requires embedding the
formulation into a constrained quadratic model (CQM).

\subsubsection{Mathematical Formalism}
The key step is to replace the continuous portfolio weights $w_{t,i}$ by a finite binary encoding. For each asset $i$, time index $t$, and bit position $q$, binary variables
\begin{equation}
x_{t,i,q} \in \{0,1\}, \qquad q=0,\dots,N_{\mathrm{enc}}-1,
\end{equation}
are introduced. Defining
\begin{equation}
M = 2^{N_{\mathrm{enc}}}-1, \qquad \Delta_i = u_i-l_i,
\end{equation}
the corresponding portfolio weight is decoded as
\begin{equation}
w_{t,i}(x) = l_i + \frac{\Delta_i}{M}\sum_{q=0}^{N_{\mathrm{enc}}-1} 2^q x_{t,i,q},
\end{equation}
where $w_{t,i}$ is defined as the portfolio weight of asset $i$, following Eq.~\eqref{eq:weights}.

This formulation respects the bounds $l_i \le w_{t,i} \le u_i$ introduced previously and, thus, preserves asset-level feasibility by construction.

Following \cite{dynamic-portfolio-optimization,Rosenberg:2015fem}, we defined the dynamic portfolio optimization function as a combination of a normalized return term $\mathbf{R}^{\text{QUBO}}$, a normalized variance penalty $\mathbf{Vol}^{\text{QUBO}}$, a transaction-cost term $\mathbf{C}^{\text{QUBO}}$, and
a budget regularizer $\mathbf{B}^{\text{QUBO}}$, defined in the following. 

Let the boundary returns $R_{\text{port}}^{\max}(t)$ and $R_{\text{port}}^{\min}(t)$, and the maximum portfolio volatility $\mathbf{Vol}^{\max}_{\text{port}}$, be defined as:
\begin{align*}
    R_{\text{port}}^{\max}(t) &= \max_{w_{t,i} \in \mathcal{W}} \sum_{i=1}^{N} w_{t,i} R_i(t,t+1), \\
    R_{\text{port}}^{\min}(t) &= \min_{w_{t,i} \in \mathcal{W}} \sum_{i=1}^{N} w_{t,i} R_i(t,t+1), \quad \text{and} \\
    \mathbf{Vol}^{\max}_{\text{port}} &= \max_{t, w_{t,i} \in \mathcal{W}} \mathbf{Vol}^{\max}_{t, \text{port}} \\
    &= \max_{t, w_{t,i} \in \mathcal{W}} \sqrt{\sum_{i=1}^{N} \sum_{j=1}^{N} w_{t,i} \, \hat{\Sigma}_{ij}(t) \, w_{t,j}}.
\end{align*}
Then, the normalized return and variance penalty terms are given by:

\begin{equation}
\begin{aligned}
    \mathbf{R}^{\text{QUBO}}_t & = \frac{\sum_{i=1}^{N} w_{t,i} R_i(t,t+1) - R_{port}^{\min}(t)}{\xi_t}
\end{aligned}
\end{equation}

\begin{equation}
\begin{aligned}
    \mathbf{Vol}^{\text{QUBO}}_t & = \frac{\sqrt{\sum_{i=1}^{N} \sum_{j=1}^{N} w_{t,i}\,\hat{\Sigma}_{ij}(t)\,w_{t,j}}}{\mathbf{Vol}_{port}^{\max}}
\end{aligned}
\end{equation}

where  $\xi_t = R_{port}^{\max}(t)-R_{port}^{\min}(t)$ denotes the range of portfolio returns at times $t$.

The transaction component deserves special attention because it is the only term that couples adjacent rebalancing dates. In the underlying financial problem, transaction costs are naturally proportional to turnover and therefore scale as $f_i |\Delta w_{t,i}|$, where
\begin{equation}
\Delta w_{t,i} = w_{t,i} - w_{t,i}^{\mathrm{drift}}.
\end{equation}
However, a QUBO model only admits quadratic terms in the binary variables. For that reason, the absolute transaction-cost contribution is approximated by a quadratic approximation over the feasible turnover interval. Using the turnover-bound notation introduced in the constraints section, $\tau_i$ denotes the maximum admissible weight change for asset $i$. The implementation then uses the effective coefficient
\begin{equation}
\tilde f_{t,i} =
\frac{2^{1/3} f_i}
{\tau_i\xi_{t+1}},
\end{equation}
so that the normalized transaction term becomes
\begin{equation}
\begin{aligned}
\mathbf{C}^{\text{QUBO}}_t =
& \sum_{i=1}^{N}
\tilde f_{t,i}
\Big(w_{t+1,i}-w_{t,i}^{\mathrm{drift}}\Big)^2.
\end{aligned}
\end{equation}

Note that the quadratic transaction term used in the quantum formulation approximates the linear transaction costs used in the classical benchmark, which may introduce slight differences in effective penalization.

The factor $\frac{2^{1/3}}{\tau_i}$ comes from the calibration of the quadratic surrogate to the absolute-value function over that interval. More precisely, if one approximates $|x|$ on $[0,\tau_i]$ by a function of the form $a x^2$, the choice

\begin{equation}
a = \frac{2^{1/3}}{\tau_i}~,
\end{equation}

minimizes the integrated absolute approximation error. 

\begin{equation}
    \epsilon = \int_{-\tau_i}^{\tau_i} ||x|-ax^2|
\end{equation}

This adjustment improves the fidelity of the quadratic transaction term relative to the original proportional trading-cost model and prevents the turnover penalty from becoming artificially weak when the admissible changes are small.

\begin{comment}
    After substituting the binary encoding of the weights, this term becomes
    \begin{equation}
    \begin{aligned}
    & \lambda \sum_{t=0}^{T-2}\sum_{i=1}^{N}
    \tilde f_{t,i} \\
    & \quad \times \left(
    \alpha_{t,i}
    + \beta_i \sum_q 2^q x_{t+1,i,q}
    - \gamma^{\mathrm{tr}}_{t,i} \sum_q 2^q x_{t,i,q}
    \right)^2,
    \end{aligned}
    \end{equation}
    with
    \begin{equation}
    \alpha_{t,i} = l_i\big(1-e^{\mu_{t,i}}\big),
    \qquad
    \beta_i = \frac{\Delta_i}{K},
    \qquad
    \gamma^{\mathrm{tr}}_{t,i} = \frac{\Delta_i e^{\mu_{t,i}}}{K}.
    \end{equation}
    This expansion makes explicit how the intertemporal transaction term decomposes into constant, linear, and quadratic binary contributions while maintaining the quadratic structure required by the annealing-based solver.  
\end{comment}

Lastly, budget's soft constraint could be expressed as

\begin{equation} \label{budget_term}
\begin{aligned}
    \mathbf{B}_t^{\text{QUBO}} & = \left(\sum_{i=1}^{N} w_{t,i}-1\right)^2~.
\end{aligned}
\end{equation}

Given this structure, the objective can be written as
\begin{equation}
\begin{aligned}
f_{\mathrm{QUBO}}(w) = & \sum_{t=0}^{T-1} \Bigg[
-\mathbf{R}^{\text{QUBO}}_t + \gamma \mathbf{Vol}^{\text{QUBO}}_t + \rho \mathbf{B}^{\text{QUBO}}_t
\Bigg] \\
&+ \lambda \sum_{t=0}^{T-2} \mathbf{C}^{\text{QUBO}}_t~.
\end{aligned}
\end{equation}

The quantities $\xi_t$, and $V_{port}^{\max}$ are normalization constants used to keep the return and risk contributions comparable. The parameter $\gamma$ controls risk aversion, $\lambda$ controls the strength of transaction-cost penalization, and $\rho$ penalizes deviations from the budget condition. In the paper, the normalization is applied in order to make the different components of the objective play on equal footing and to avoid one term dominating the others purely because of scale.

After substituting the binary expansion of each weight into the objective and converting all linear terms to quadratic terms by exploiting the property of binary variables $x_i^2=x_i$, it is possible to show that the multi-period problem takes the standard QUBO form
\begin{equation}
f_{\mathrm{QUBO}}(x) = x^{\top}Qx + b,
\end{equation}
where $Q$ is the matrix of binary couplings and $b$ is a constant offset. The entries of $Q$ are assembled from the return contribution, the variance contribution, the budget penalty, and the expanded transaction-cost contribution. 

In practice, the optimization problem considered in this work also contains feasibility requirements, introduced in Section~\ref{Constraints}, that are more naturally enforced as explicit constraints than as soft penalties alone. For that reason, the QUBO objective is embedded into a constrained quadratic model (CQM), where asset-level turnover limits, exposure bounds, and class-level restrictions are imposed directly. This model is then solved using D-Wave's LeapHybridCQMSampler, corresponding to the hybrid quantum-classical solver for constrained quadratic models available.

\subsubsection{Implementation}

To make the full backtesting horizon compatible with the solver capacity, the multi-period optimization is not encoded as a single problem over all 418 rebalancing dates. Instead, the horizon is decomposed into yearly blocks, each containing 26 biweekly rebalancing dates. Within each block, the complete sequence of portfolio weights is optimized jointly, so the solver captures the intertemporal effect of transaction costs and turnover constraints over that year. Once a yearly block is solved, its final portfolio allocation is used as the initial condition for the next block, ensuring continuity across years and preserving the path-dependent rebalancing constraints. Thus, the approach combines global optimization within each yearly window with a sequential rollout across the full historical period.

Given the hardware characteristics of the LeapHybridCQMSampler  and the given scenario to optimize, $\sim 20$ years, we defined the following workflow in order to solve the problem: 

\begin{enumerate}
    \item Define rebalancing frequency, in this case, every two weeks.
    \item Encode the problem for a whole year, 26 rebalancing dates.
    \item Solve the problem and save the resulting optimal weights.
    \item Repeat this process for the next year using previous period optimal weights as starting weights in order to satisfy transactions limit constraints between periods.
    \item Do this iteratively for all years.
\end{enumerate}

With this workflow we can efficiently solve arbitrary large time windows without reaching hardware limitations.

The methodology described in this section provides two main advantages relative to the classical benchmarks introduced previously:
\begin{itemize}
    \item The resulting portfolio weights constitute a globally optimized sequence $w_t \in \mathbb{R}^N$ over the rebalancing dates $t$ included in the optimization, in our case one whole year. In contrast, classical approaches typically compute locally optimal allocations at each rebalancing step and connect them through transaction-cost restrictions.
    \item Classical methods such as CVaR optimization generally require a distributional assumption on asset returns, which in our setting is taken to be Gaussian, in order to derive the optimal portfolio weights. The proposed approach does not impose any return-distribution assumption, and therefore avoids this source of approximation.
\end{itemize}
For additional details, see Table \ref{tab:hybrid_solver_limits}.
\section{Results}
This section presents the comparison of the proposed quantum-classical optimization and the two classical benchmarks, that is, a rebalanced buy-and-hold allocation and a CVaR-minimizing strategy.

The analysis is divided into two complementary perspectives. 
First, we analyze the computational characteristics of the
hybrid quantum-classical optimization workflow. 
Second, we evaluate the economic quality of the resulting portfolios through return, volatility, and
Sharpe ratio under different transaction-cost regimes.

Furthermore, hardware-level results and business-oriented metrics are reported separately to distinguish economic performance from computational performance. 
This separation is important because strong financial performance does not necessarily imply computational advantage, and computational advantage does not automatically translate into superior investment performance.

%Lastly, any discussion of hardware performance should avoid unsupported claims of quantum advantage unless such a conclusion is backed by direct and reproducible benchmarking evidence against classical discrete optimization methods under comparable conditions.

\subsection{Hardware Performance}

In this part, we discuss the computational aspects of the proposed framework.
These include the size of the binary formulation (i.e., the number of binary variables required by the QUBO model), the role of quadratic couplings, the handling of constraints through the CQM layer, and the operational characteristics of the hybrid quantum-classical solver.

The yearly decomposition strategy described above was adopted to remain within the practical capacity limits of the CQM and hybrid solver summarized in Table \ref{tab:hybrid_solver_limits}. This decomposition is required because the complete formulation exceeds the number of binary variables supported within a single CQM instance.

Since the main contribution of this work lies in the practical financial applicability of the proposed optimization framework, the hardware-related analysis focuses on operationally relevant quantities such as total wall-clock time, which directly reflects the effective time-to-solution in realistic portfolio-management scenarios.

\subsubsection{Execution time}

Specifically, the reported wall-clock time corresponds to the total time required by each method to complete the full portfolio optimization over the entire backtesting horizon (2003-01-06 to 2018-12-31), comprising 418 biweekly rebalancing dates.

\paragraph{Classical setup}
At each of the 418 rebalancing dates, the linear program in Eq. \eqref{eq:cvar-model} is solved using $S = 150{,}000$ Monte Carlo scenarios drawn from the parametric Gaussian generator described in Section~IV. The cumulative wall-clock time for the full rolling-horizon backtest amounts to 4~hours and 15~minutes.
The CVaR optimization was executed on a MacBook Pro running macOS Tahoe Version~26.4.1, equipped with an Apple M2 chip (8-core CPU) and 16~GB of unified memory.

\paragraph{Quantum setup}
The optimization problem was solved using D-Wave's LeapHybridCQMSampler (version~1.13), a cloud-based hybrid quantum-classical solver designed for constrained quadratic models (CQMs) \cite{dwave-hybrid-solver-2024}. The solver integrates classical optimization techniques with quantum annealing resources provided by D-Wave's Advantage quantum processing unit (QPU), which comprises more than 5,000 qubits arranged in a Pegasus topology with 15-way qubit connectivity  \cite{dwave-advantage-system}. The hybrid workflow decomposes the problem into subproblems that are distributed between classical hardware and the QPU. As described in Section~IV, the full 16-year backtest is decomposed into yearly batches of 26 rebalancing dates each in order to remain within the solver capacity limits summarized in Table~\ref{tab:hybrid_solver_limits}. The cumulative wall-clock time for solving the entire optimization period, including network latency and QPU access, amounts to 35~minutes.

Table~\ref{tab:hardware_comparison} presents the computational environment specifications and the corresponding timing results obtained for both optimization approaches.
The hybrid quantum-classical workflow achieves a wall-clock reduction factor of approximately $7.3\times$ relative to the classical CVaR implementation considered in this study.
This comparison should be interpreted with nuance: the two methods differ not only in hardware but also in algorithmic structure. The CVaR approach solves 418 independent linear programs sequentially, each involving $150{,}000$ scenarios and the full set of portfolio constraints, whereas the quantum approach solves the problem for 26 rebalancing periods simultaneously with a single CQM instance per yearly batch, jointly optimizing multiple rebalancing periods within each yearly batch. 

Moreover, the hybrid solver offloads computation to a remote cloud infrastructure with dedicated high-performance classical resources in addition to the QPU, whereas the classical benchmark runs entirely on a consumer-grade laptop.
Consequently, the observed timing difference reflects differences in problem formulation, hardware architecture, and execution model, rather than a pure quantum-versus-classical comparison. A rigorous claim of quantum computational advantage would require benchmarking against equivalent discrete optimization methods on comparable hardware, which remains outside the scope of this work. Nevertheless, the result demonstrates that the hybrid quantum-classical approach is practically viable for real-world portfolio optimization at this problem scale, delivering competitive solutions in substantially less wall-clock time.

\begin{table}[ht]
\caption{Hardware and Execution Time Comparison}
\label{tab:hardware_comparison}
\centering
\renewcommand{\arraystretch}{1.15}
\resizebox{\columnwidth}{!}{%
\begin{tabular}{lll}
\hline
 & \textbf{Classical (CVaR)} & \textbf{Quantum (CQM)} \\
\hline
Processor & Apple M2 (8-core) & D-Wave Advantage  \\
Resources & 16 GB unified & $>$5000 qubits \\
OS / Platform & macOS Tahoe 26.4.1 & D-Wave Leap (cloud) \\
Solver & CVXPy & LeapHybridCQMSampler v1.13 \\
Optimization calls & 418 (sequential LP) & 16 (yearly CQM batches) \\
Total wall-clock time & 4 h 15 min & 35 min\\
Speedup factor & --- & $\approx 7.3\times$ \\
\hline
\end{tabular}%
}
\end{table}

\subsection{Economic Performance}

This section presents the empirical evaluation of the proposed framework. We evaluate the economic quality of the resulting portfolios through \emph{return}, \emph{volatility}, and \emph{Sharpe ratio} under different transaction-cost regimes. In this way, the study distinguishes not only between absolute return and risk levels but the balance between them.

From a business perspective, the key question is whether the additional flexibility of the quantum formulation translates into improved risk-adjusted performance when transaction costs are incorporated. Conversely, classical baselines provide a benchmark for evaluating whether the additional modeling complexity of the quantum formulation leads to economically meaningful improvements.

\begin{figure}
    \centering
    \includegraphics[width=0.75\linewidth]{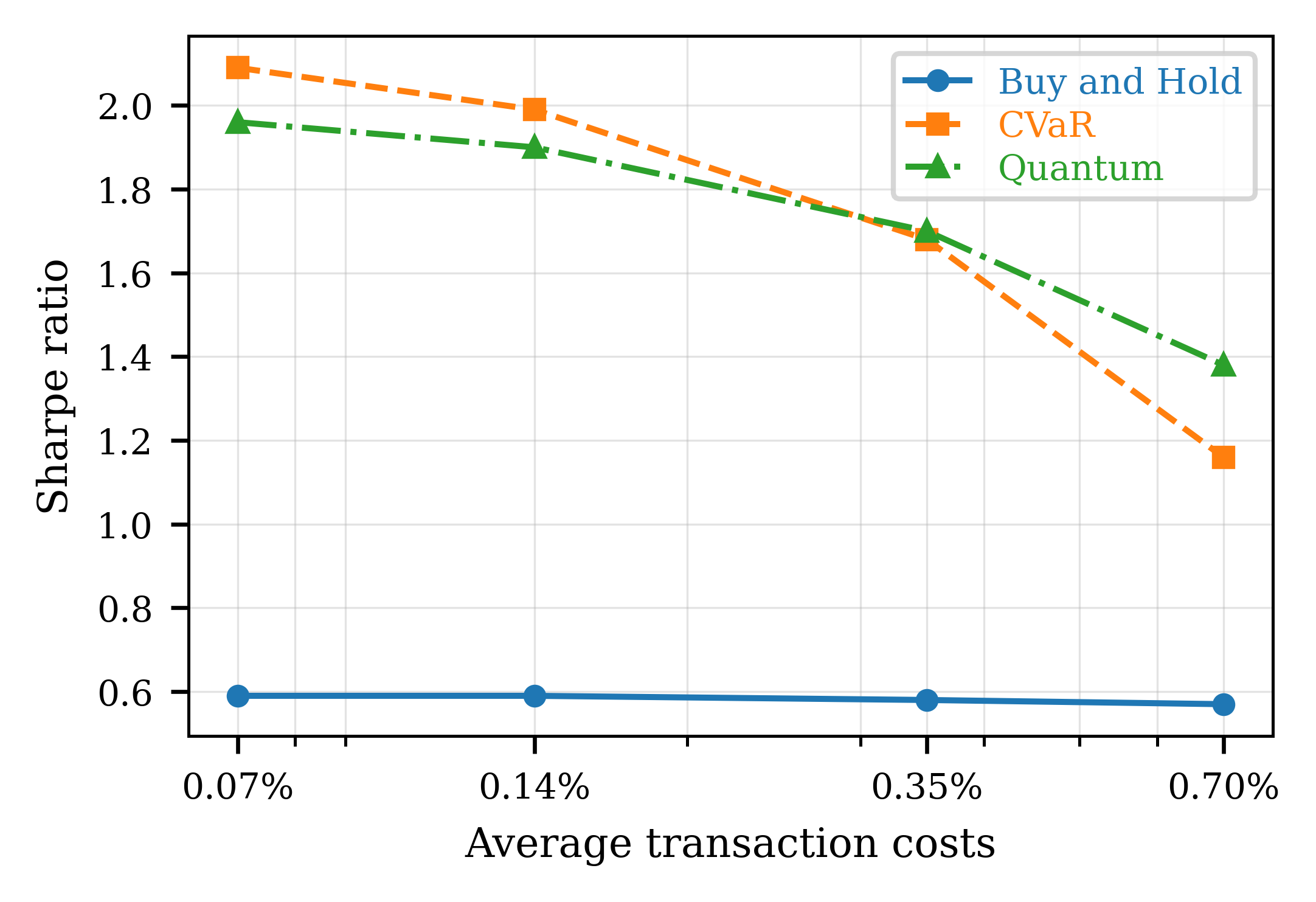}
    \caption{Average Sharpe ratio across different transaction-cost levels. Quantum-derived portfolios become
competitive with CVaR above $35\, bp$.}
    \label{fig:Sharpe ratio Comparison}
\end{figure}

The historical rolling-window backtest over the 2003--2018 period indicates that classical and quantum methods achieve relatively similar performance under low transaction-cost regimes (see Figure~\ref{fig:Sharpe ratio Comparison}). Under low-cost regimes—where fees average approximately 7 basis points— CVaR achieves a higher Sharpe ratio than the quantum-derived portfolio (2.09 for CVaR versus 1.96). As transaction costs rise, however, the performance gap narrows progressively until both methods reach approximate parity. Beyond a break-even point of roughly 35 basis points, the quantum-derived portfolios begin to yield a more competitive risk-adjusted performance (Sharpe ratio of 1.70 versus 1.68 for CVaR at 35 basis points, and 1.38 versus 1.16 for CVaR at 70 basis points).
This suggests that the global multi-period optimization of the quantum strategy becomes competitive with CVaR as transaction costs increase.

% Under low-friction regimes (average fees of 7 basis points), CVaR achieves a higher Sharpe ratio. As costs rise, the gap narrows until the two strategies reach equal performance near 35 basis points, beyond which the quantum-derived portfolio delivers a superior risk-adjusted performance. This crossover threshold lies well within realistic market conditions---for reference, UK stamp duty alone amounts to 50 basis points on equity purchases---confirming that the observed advantage is economically relevant rather than an artifact of extreme parameterization.
 Notably, this threshold is well within realistic market conditions: for reference, the stamp duty levied on equity purchases in the United Kingdom amounts to 50 basis points. Consequently, the 35 basis-point level at which quantum-derived portfolios begin to outperform represents a realistic and economically relevant scenario rather than an extreme edge case.
More specifically, we can see:
\begin{itemize}
    \item In Table \ref{tab:return_comparison}, that the quantum-derived portfolio achieves the highest annualized return. Focusing on the break-even case of 35 basis points, $10.57\%$ vs $8.06\%$ for the best classical approach (CVaR).
    \item In Table \ref{tab:vol_comparison}, that the quantum-derived portfolio ranks second in terms of risk, with an annual volatility of $5.58\%$ vs $4.18\%$ of CVaR strategy (the best for risk-averse investors) or $8.14\%$ of buy-and-hold strategy (the riskiest). 
    \item and finally, in Table \ref{tab:sharpe_comparison} and in Figure \ref{fig:Sharpe ratio Comparison}, that the quantum-derived portfolio achieves a higher Sharpe ratio than all classical methods when transaction costs are above $35$ basis points.
\end{itemize}

% TODO: Insert exact tables and figures once the notebook and spreadsheet are re-uploaded.

%    TODO: Insert exact hardware and solver diagnostics once available.

% =================================================≈
\section{Conclusion}

\begin{comment}
This paper proposes a structured comparison between quantum and classical optimization strategies for constrained investment portfolio construction. The quantum formulation relies on a binary encoding of portfolio weights, a QUBO representation of the economic objective, and a constrained quadratic model for the exact enforcement of feasibility conditions. The classical side of the comparison provides economically interpretable baselines against which the quantum strategy can be assessed.

The resulting framework supports a dual evaluation. On the one hand, it enables the analysis of business-level portfolio performance under realistic rebalancing frictions. On the other hand, it makes it possible to study the computational behavior of the underlying optimization machinery. This two-layer perspective is essential for a balanced assessment of quantum optimization in finance, where practical relevance depends simultaneously on economic validity and computational tractability.

\end{comment}

% =================================================

This paper presents an empirical comparison between a quantum strategy for dynamic portfolio optimization and two classical benchmarks, a rebalanced buy-and-hold allocation and a CVaR-minimizing strategy, under realistic trading constraints. The study considers a diversified investment universe of 21 global assets spanning equities, government bonds, credit, foreign exchange, and cash, subject to asset-level, class-level, and turnover constraints over 418 biweekly rebalancing periods (2003--2018).

The quantum formulation encodes portfolio weights into binary variables and constructs a constrained quadratic model (CQM) solved by D-Wave's hybrid quantum-classical sampler. 
In contrast to the classical CVaR benchmark, which optimizes sequentially at each rebalancing date under a Gaussian return assumption while using previous portfolio weights as the initial condition, the quantum approach jointly optimizes the full trajectory over yearly windows without imposing a parametric distributional model on asset returns.

The key result is that the quantum-derived strategy improves monotonically as transaction costs increase, eventually overtaking the CVaR benchmark.
Under low-transaction-costs regimes (average fees of 7 basis points), CVaR achieves a higher Sharpe ratio; however, as costs rise, the performance gap gradually narrows until both strategies converge near 35 basis points. Beyond this threshold, the quantum-derived portfolio achieves a higher Sharpe ratio than the CVaR benchmark.

Importantly, this crossover occurs well within realistic market conditions —for reference, UK stamp duty alone amounts to 50 basis points on equity purchases— confirming the practical relevance of the observed advantage.

Additionally, the quantum strategy achieves the highest annualized return across all fee scenarios, while maintaining intermediate volatility.
This pattern further suggests that the sequential nature of the CVaR optimization becomes progressively less effective at managing cumulative rebalancing frictions than the multi-period trajectory optimization enabled by the quantum formulation.

Several limitations must be acknowledged. 
First, the QUBO formulation approximates the linear transaction-cost function using a quadratic surrogate, introducing a modeling error relative to the true transaction costs. 
Second, the problem was decomposed into yearly subproblems to ensure computational feasibility on current quantum hardware, yielding a tractable, albeit non-exact, approximation to the full multi-year optimization problem.

Despite these caveats, the results show that quantum-derived dynamic portfolio optimization becomes increasingly competitive as realistic transaction costs are incorporated, with the break-even occurring at transaction-cost levels commonly observed in institutional practice.
These findings suggest that quantum methods may be particularly well suited to constrained, multi-period portfolio optimization problems where rebalancing costs play a significant role.
\section{Future Work}

Several research extensions arise naturally from this work.
First, the sensitivity of the quantum solution to the discretization depth, penalty calibration, and normalization scheme should be studied systematically.
Second, future work may compare the hybrid constrained quadratic model approach with alternative classical discrete optimization techniques, such as mixed-integer quadratic programming or metaheuristic search methods, under identical problem instances and conditions. 
Third, the portfolio design can be extended to encompass richer risk measures, alternative transaction-cost models, and scenario-based uncertainty representations.
Finally, a more exhaustive computational study is essential before any claim regarding potential quantum advantage in portfolio optimization can be responsibly formulated.

Additionally, while the empirical results suggest that the quantum-derived optimal portfolios exhibit favorable risk-adjusted performance, the current reliance on offline historical data limits their direct use in real-time industrial settings.

To bridge this gap, a promising direction involves utilizing the quantum-optimized allocations as training targets for a supervised learning surrogate model.
Once trained on relevant macro and microeconomic market variables, such a model could generate real-time allocation recommendations at minimal operational latency, while potentially uncovering interpretable market regimes that inform adaptive investment strategies.
This line of research is currently under active development and constitutes the immediate continuation of the present work.

% =================================================

\section*{Acknowledgment}
The authors thank BBVA Corporate and Investment Banking and Asset Management for helpful discussions and technical support.\\
%\myCommentESM{esto que ponéis aquí, qué es?}\myCommentABG{Son agradecimientos al proyecto nacional en el que estamos Guillermo y yo. Normalmente agradecemos a todos los proyectos relacionados que tenemos vigentes (éste tiene un componente de Cuántica muy alto). Si no os parece bien, lo quitamos.} This work was partially supported by grant  PID2024-158203OB-I00 funded by MICIU/AEI/10.13039/501100011033 and by ERDF/EU. \myCommentABG{Deberíamos incluir agradecimientos a Q-MIND?} \myCommentESM{Hemos pedido el aknowledgement a GMV, añadido debajo}\\
This work was developed within the framework of the Q-MIND consortium, co-funded by the European Union through the European Regional Development Fund (ERDF), under the Programa de la Comunidad de Madrid for the 2021–2027 period promoted by the Consejería de Educación, Ciencia e Universidades.

% =================================================

\bibliographystyle{IEEEtran}
\bibliography{references}

% =================================================

\clearpage
\appendices

\section{Results Tables}
In this appendix we present the key financial performance metrics used to evaluate the optimized portfolios:

\begin{itemize}
    \item Annualized return
    \item Annualized volatility
    \item Annualized Sharpe ratio
\end{itemize}

These metrics are reported under four transaction-cost regimes. The column ``Avg.\ TC'' denotes the portfolio-weighted average two-way transaction cost, expressed as a percentage of portfolio value (e.g., $0.07\%$ corresponds to 7 basis points). Volatility is approximately invariant across cost regimes; transaction costs affect only net returns and, consequently, the Sharpe ratio.

\begin{table}[H]
\caption{Return comparison under different transaction cost levels}
\label{tab:return_comparison}
\centering
\renewcommand{\arraystretch}{1.10}
\resizebox{\columnwidth}{!}{%
\begin{tabular}{cccc}
\hline
Avg. TC & Buy \& Hold & CVaR & Quantum \\
\hline
0.07\% & 5.86\% & 9.82\% & 12.00\% \\
0.14\% & 5.84\% & 9.38\% & 11.64\% \\
0.35\% & 5.79\% & 8.06\% & 10.57\% \\
0.70\% & 5.69\% & 5.87\% & 8.78\% \\
\hline
\end{tabular}%
}
\end{table}

\begin{table}[H]
\caption{Volatility comparison under different transaction cost levels}
\label{tab:vol_comparison}
\centering
\renewcommand{\arraystretch}{1.10}
\resizebox{\columnwidth}{!}{%
\begin{tabular}{cccc}
\hline
Avg. TC & Buy \& Hold & CVaR & Quantum \\
\hline
0.07\% & 8.14\% & 4.19\% & 5.58\% \\
0.14\% & 8.14\% & 4.19\% & 5.58\% \\
0.35\% & 8.14\% & 4.18\% & 5.58\% \\
0.70\% & 8.14\% & 4.18\% & 5.58\% \\
\hline
\end{tabular}%
}
\end{table}

\begin{table}[H]
\caption{Sharpe ratio comparison under different transaction cost levels}
\label{tab:sharpe_comparison}
\centering
\renewcommand{\arraystretch}{1.10}
\resizebox{\columnwidth}{!}{%
\begin{tabular}{cccc}
\hline
Avg. TC & Buy \& Hold & CVaR & Quantum \\
\hline
0.07\% & 0.59 & 2.09 & 1.96 \\
0.14\% & 0.59 & 1.99 & 1.90 \\
0.35\% & 0.58 & 1.68 & 1.70 \\
0.70\% & 0.57 & 1.16 & 1.38 \\
\hline
\end{tabular}%
}
\end{table}

\section{Hardware Tech Specs}

In this appendix you can find the detailed technical specifications of the Quantum Annealer utilized for the development of this work.

\begin{table}[H]
\caption{Constrained Hybrid solver technical specs.}
\label{tab:hybrid_solver_limits}
\centering
\begin{tabular}{ll}
\hline
\textbf{Parameter} & \textbf{Value} \\
\hline
Category & Hybrid \\
Maximum number of biases & $2{,}000{,}000{,}000$ \\
Maximum number of constraints & $100{,}000$ \\
Maximum number of quadratic variables & $100{,}000{,}000$ \\
Maximum number of variables & $500{,}000$ \\
Supported problem types & CQM \\
Version & $1.13$ \\
\hline
\end{tabular}
\end{table}

\section{Asset Universe}\label{Assets} 

In this appendix you can find all the details about the assets used in this work.

The column headers are defined as follows: \textbf{Min} and \textbf{Max} denote the lower and upper bounds on individual asset weights; \textbf{Chg.}\ denotes the maximum absolute change in weight permitted at each biweekly rebalancing date (i.e., $|w_{t,i} - w_{t-1,i}| \le \text{Chg.}$); \textbf{$f$} denotes the two-way transaction cost in basis points; and \textbf{B\&H} denotes the fixed target allocation used in the buy-and-hold benchmark.

All indices are sourced from Bloomberg and represent total-return series denominated in EUR.
 
\begin{table}[H]
\centering
\caption{Equity asset characteristics and constraints}
\label{tab:equities}
\scriptsize
\setlength{\tabcolsep}{2.5pt}
\renewcommand{\arraystretch}{0.96}
\begin{threeparttable}
\begin{tabular}{@{}>{\raggedright\arraybackslash}p{0.46\columnwidth}ccccc@{}}
\toprule
Asset / ticker & Min & Max & Chg. & $f$ & B\&H \\
\midrule
US eq. / {GDDUUS} & 5\% & 45\% & 8\% & 3 & 25\% \\
Europe eq. / {GDDLE15} & 7\% & 23\% & 3\% & 3 & 15\% \\
Japan eq. / {GDDLJN} & 0\% & 6\% & 2\% & 3 & 3\% \\
EM eq. / {M2EF} & 2\% & 12\% & 2\% & 3 & 7\% \\
\bottomrule
\end{tabular}
\begin{tablenotes}[flushleft]
\footnotesize
\item Notes: Portfolio weights are reported in Min, Max, Chg., and B\&H. $f$ denotes two-way transaction costs in basis points.
\end{tablenotes}
\end{threeparttable}
\end{table}

\begin{table}[H]
\centering
\caption{Government-bond asset characteristics and constraints}
\label{tab:government-bonds}
\scriptsize
\setlength{\tabcolsep}{2.2pt}
\renewcommand{\arraystretch}{0.95}
\begin{threeparttable}
\begin{tabular}{@{}>{\raggedright\arraybackslash}p{0.49\columnwidth}ccccc@{}}
\toprule
Asset / ticker & Min & Max & Chg. & $f$ & B\&H \\
\midrule
US gov. 7--10y / {G4O2} & 0\% & 10\% & 3\% & 4.5 & 4\% \\
US gov. 5--7y / {G3O2} & 0\% & 10\% & 4\% & 3.75 & 3\% \\
US gov. 1--3y / {G1O2} & 0\% & 10\% & 4\% & 1.5 & 3\% \\
Euro gov. 7--10y / {EG04} & 0\% & 10\% & 3\% & 5.25 & 4\% \\
Euro gov. 5--7y / {EG03} & 0\% & 10\% & 4\% & 4.38 & 3\% \\
Euro gov. 1--3y / {EG01} & 0\% & 10\% & 3\% & 1.75 & 3\% \\
EM gov. HC / {EMGB} & 0\% & 10\% & 3\% & 10 & 5\% \\
\bottomrule
\end{tabular}
\begin{tablenotes}[flushleft]
\footnotesize
\item Notes: Portfolio weights are reported in Min, Max, Chg., and B\&H. $f$ denotes two-way transaction costs in basis points. HC stands for hard currency.
\end{tablenotes}
\end{threeparttable}
\end{table}

\begin{table}[H]
\centering
\caption{Credit asset characteristics and constraints}
\label{tab:credit}
\scriptsize
\setlength{\tabcolsep}{2.5pt}
\renewcommand{\arraystretch}{0.96}
\begin{threeparttable}
\begin{tabular}{@{}>{\raggedright\arraybackslash}p{0.47\columnwidth}ccccc@{}}
\toprule
Asset / ticker & Min & Max & Chg. & $f$ & B\&H \\
\midrule
US credit / {C0A0} & 0\% & 18\% & 4\% & 10 & 9\% \\
EUR credit / {ERL0} & 0\% & 18\% & 4\% & 10 & 9\% \\
Pan-Euro HY / {HE00} & 0\% & 10\% & 3\% & 18 & 3.5\% \\
US HY / {H0A0} & 0\% & 10\% & 3\% & 15 & 3.5\% \\
\bottomrule
\end{tabular}
\begin{tablenotes}[flushleft]
\footnotesize
\item Notes: Portfolio weights are reported in Min, Max, Chg., and B\&H. $f$ denotes two-way transaction costs in basis points. HY stands for high yield.
\end{tablenotes}
\end{threeparttable}
\end{table}

\begin{table}[H]
\centering
\caption{Cash asset characteristics and constraints}
\label{tab:cash}
\scriptsize
\setlength{\tabcolsep}{2.5pt}
\renewcommand{\arraystretch}{0.96}
\begin{threeparttable}
\begin{tabular}{@{}>{\raggedright\arraybackslash}p{0.56\columnwidth}ccccc@{}}
\toprule
Asset / ticker & Min & Max & Chg. & $f$ & B\&H \\
\midrule
EUR cash o/n / {L0EC} & 0\% & 60\% & 11\% & 0 & 0\% \\
\bottomrule
\end{tabular}
\begin{tablenotes}[flushleft]
\footnotesize
\item Notes: Portfolio weights are reported in Min, Max, Chg., and B\&H. $f$ denotes two-way transaction costs in basis points.
\end{tablenotes}
\end{threeparttable}
\end{table}

\begin{table}[H]
\centering
\caption{FX-forward asset characteristics and constraints}
\label{tab:fx-forwards}
\scriptsize
\setlength{\tabcolsep}{2.5pt}
\renewcommand{\arraystretch}{0.96}
\begin{threeparttable}
\begin{tabular}{@{}>{\raggedright\arraybackslash}p{0.58\columnwidth}cccc@{}}
\toprule
Asset / ticker & Min & Max & Chg. & $f$ \\
\midrule
EUR/USD 45d / {BBLIXEUE} & -20\% & 20\% & 4\% & 0 \\
USD/JPY 45d / {BBLIXJPE} & -10\% & 10\% & 2\% & 0 \\
USD/CAD 45d / {BBLIXCAE} & -10\% & 10\% & 2\% & 0 \\
USD/AUD 45d / {BBLIXAUE} & -10\% & 10\% & 2\% & 0 \\
USD/MXN 45d / {BBLIXMXE} & -10\% & 10\% & 2\% & 0 \\
\bottomrule
\end{tabular}
\begin{tablenotes}[flushleft]
\footnotesize
\item Notes: Portfolio weights are reported in Min, Max, and Chg. $f$ denotes two-way transaction costs in basis points.
\end{tablenotes}
\end{threeparttable}
\end{table}

\begin{table}[H]
\centering
\caption{Aggregate limits by asset class}
\label{tab:aggregate-limits}
\footnotesize
\setlength{\tabcolsep}{2.5pt}
\renewcommand{\arraystretch}{0.96}
\begin{threeparttable}
\begin{tabular}{@{}lcccc@{}}
\toprule
Class & Min & Max & Chg. & B\&H \\
\midrule
Equity & 30\% & 70\% & 8\% & 50\% \\
Gov. & 5\% & 45\% & 8\% & 25\% \\
Cred. & 5\% & 45\% & 8\% & 25\% \\
Cash & 0\% & 60\% & 12\% & 0\% \\
\bottomrule
\end{tabular}
\begin{tablenotes}[flushleft]
\footnotesize
\item Notes: Asset class weights are reported in Min, Max, and Chg. B\&H denotes the asset class weights used in the buy-and-hold portfolio.
\end{tablenotes}
\end{threeparttable}
\end{table}

\section{Annualization Convention for CVaR}
\label{app:cvarannaulization}

This appendix describes the annualization convention used for CVaR, where we use a Gaussian time-aggregation approximation. Let
\(\mathrm{CVaR}_{d,\alpha}\) denote the daily CVaR at confidence level
\(\alpha\), expressed as a positive loss measure. Under the assumption
that daily returns are independent and identically distributed with
constant mean and volatility, the annualized CVaR is computed as
\begin{equation}
    \mathrm{CVaR}_{\mathrm{ann},\alpha}
    =
    \sqrt{D}\,\mathrm{CVaR}_{d,\alpha}
    -
    (D-\sqrt{D})\,\bar r_d .
\end{equation}

This expression combines linear scaling of the mean return component
with square-root-of-time scaling of the tail-risk component. Its
derivation is provided below. The formula is exact under the i.i.d.
Gaussian additive-return assumption and is used here as an
annualization convention outside that setting.

\paragraph{Derivation}
The expression above follows from a Gaussian time-aggregation argument.
Let \(r_t\) denote the daily portfolio return and define the daily loss as
\(L_t=-r_t\). Assume that daily returns are independent and identically
distributed as
\begin{equation}
    r_t \sim \mathcal{N}(\mu_d,\sigma_d^2).
\end{equation}
Then
\begin{equation}
    L_t \sim \mathcal{N}(-\mu_d,\sigma_d^2).
\end{equation}

For a normally distributed loss variable \(L=m+sZ\), with
\(Z\sim\mathcal{N}(0,1)\), the CVaR at confidence level \(\alpha\) is
\begin{equation}
    \mathrm{CVaR}_{\alpha}(L)
    =
    m+s\kappa_{\alpha},
    \qquad
    \kappa_{\alpha}
    =
    \frac{\phi(z_{\alpha})}{1-\alpha},
    \qquad
    z_{\alpha}=\Phi^{-1}(\alpha),
\end{equation}
where \(\Phi\) and \(\phi\) are the standard normal distribution
function and density, respectively. Therefore, the daily CVaR expressed
as a positive loss measure is
\begin{equation}
    \mathrm{CVaR}_{d,\alpha}
    =
    -\mu_d+\sigma_d\kappa_{\alpha}.
    \label{eq:daily-cvar-loss}
\end{equation}

Now consider the annual additive return
\begin{equation}
    R_D=\sum_{t=1}^{D}r_t .
\end{equation}
Under the i.i.d. Gaussian assumption,
\begin{equation}
    R_D\sim\mathcal{N}(D\mu_d,D\sigma_d^2),
\end{equation}
and the corresponding annual loss \(L_D=-R_D\) satisfies
\begin{equation}
    L_D\sim\mathcal{N}(-D\mu_d,D\sigma_d^2).
\end{equation}
Hence,
\begin{align}
    \mathrm{CVaR}_{\mathrm{ann},\alpha}
    &=
    -D\mu_d+\sqrt{D}\sigma_d\kappa_{\alpha}.
\end{align}
Using Eq.~\eqref{eq:daily-cvar-loss},
\begin{equation}
    \sigma_d\kappa_{\alpha}
    =
    \mathrm{CVaR}_{d,\alpha}+\mu_d.
\end{equation}
Substituting this into the previous expression gives
\begin{align}
    \mathrm{CVaR}_{\mathrm{ann},\alpha}
    &=
    -D\mu_d
    +
    \sqrt{D}\left(
        \mathrm{CVaR}_{d,\alpha}+\mu_d
    \right) \\
    &=
    \sqrt{D}\,\mathrm{CVaR}_{d,\alpha}
    -
    (D-\sqrt{D})\mu_d.
\end{align}
In the empirical implementation, \(\mu_d\) is replaced by the sample
mean \(\bar r_d\). The result is exact under the i.i.d. Gaussian
additive-return assumption and is used as an annualization convention
outside that setting.

\end{document}